\documentclass{amsart}

\makeatletter
\renewcommand{\email}[2][]{%
  \ifx\emails\@empty\relax\else{\g@addto@macro\emails{,\space}}\fi%
  \@ifnotempty{#1}{\g@addto@macro\emails{\textrm{(#1)}\space}}%
  \g@addto@macro\emails{#2}%
}
\makeatother

\usepackage{graphicx}
\usepackage{amsaddr}
\usepackage{epstopdf, epsfig}
\usepackage{natbib}

\usepackage{amsmath}
\usepackage{color}
\usepackage{upgreek}

\def\blue{\textcolor{blue}}
\def\red{\textcolor{red}}

\def\black{\textcolor{black}}

\definecolor{darkblue}{RGB}{83,0,93}
\def\L{\mbox{------}}
\def\dashL{\mbox{-~-~-}}

\def\cdotL{\mbox{--- $\cdot$ ---}}

\def\magenta{\textcolor{magenta}}
\def\blue{\textcolor{blue}}
\def\red{\textcolor{red}}

\def\black{\textcolor{black}}

\usepackage{float}

\def\nn{\nonumber}
\def\d{\textrm d}
\def\i{\textrm i}
\def\ep{\epsilon}
\def\ba#1\ea{\begin{align}#1\end{align}}
\def\!#1\!{}

\def\p{\partial}

\def\bsa#1#2\esa{\begin{subequations}\label{#1}
\begin{align}#2\end{align} \end{subequations}}

\def\lb{\left[}
\def\rb{\right]}

\def\O{{\mathcal{O}}}

\def\bsa#1#2\esa{\begin{subequations}\label{#1}
		\begin{align}#2\end{align} \end{subequations}}
\def\ba#1\ea{\begin{align}#1\end{align}}

\numberwithin{equation}{section}
\usepackage[margin=1in]{geometry}

\begin{document}

\title[Internal Wave Energy over Seabed Corrugations]{Sensitivity of internal wave energy distribution over seabed corrugations to adjacent seabed features}


\author{F. Karimpour, \ \ A. Zareei, \ \ \& M.-R. Alam}
\address{Department of Mechanical Engineering, University of California, Berkeley}
\curraddr{}
\email{fkarimpour@berkeley.edu,\ zareei@berkeley.edu, \ reza.alam@berkeley.edu}
\thanks{}



\date{}

\dedicatory{}

\begin{abstract}
Here we show that the distribution of internal gravity waves energy over a patch of seabed corrugations strongly depends on the \textit{distance} of the patch to adjacent seafloor features. Specifically, we consider the energy distribution over a patch of seabed ripples neighbored to i. another patch of ripples, and ii. a vertical wall. Seabed undulations with dominant wavenumber twice as large as overpassing internal waves reflect back part of the energy of the internal waves (Bragg reflection), let the rest of the energy to transmit or to be transferred to higher and lower modes. In the presence of a neighboring topography on the downstream side, the transmitted energy from the patch may reflect back, e.g. partially if the downstream topography is another set of seabed ripples, or fully if it is a vertical wall. The reflected wave from downstream topography is again reflected back by the patch of ripples through the same mechanism. This consecutive reflection goes on indefinitely leading to a complex interaction pattern including constructive and destructive interference of multiply reflected waves as well as an interplay between higher modes internal waves resonated over the topography. We show here that when steady state is reached both the qualitative and quantitative behavior of energy distribution over the patch is a strong function of the \textit{distance} between the patch and the downstream topography. As a result, for instance, the local energy density in the water column can become an order of magnitude larger in certain areas merely based on where the downstream topography is. This may result in the formation of steep waves in specific areas of the ocean, leading to breaking and enhanced mixing. At a right distance, the wall or the second patch may also result in a complete disappearance of the trace of the seabed undulations on the upstream and the downstream wave field.  
\end{abstract}

\maketitle

\section{Introduction}\label{sec:intro}

It is known that if an internal wave travels over a patch of corrugated seabed with twice as large wavenumber, then the energy of the incident internal wave is partially reflected, partially transferred to other modes (higher and lower), and the rest keeps traveling, i.e. transmitted to, the downstream \cite[e.g.][]{BUEHLER[2011],couston2016}. Reflection of waves as they travel in a periodic medium of double the wavelength is commonly known as Bragg reflection (or resonance). The phenomenon was first discovered in the context of electromagnetic waves in early 20th century \cite[][]{bragg1913}, and since then has been observed, elucidated and reported extensively in many other physical systems such as in solid state physics, optics, and  acoustics \cite[e.g.][]{Fermi1947,Kryuchkyan2011}, as well as in water waves \cite[e.g.][]{Mei1985,Alam2014e,alam2010aaa,Alam2009a,Alam2009b}.


Of interest of this manuscript is the dynamics of internal waves over a patch of seabed corrugations in the presence of a reflecting object downstream of the patch. This interest is motivated by several observations of enhanced (by orders of magnitude) and intense mixing over rough topographies of the oceans and the claimed attribution of these observations to internal waves breaking \cite[e.g.][]{Ledwell2000,Naveira2004}, as well as reports of strong internal waves generation over undular seabed \cite[e.g.][]{Kranenburg1991,Pietrzak1991,Pietrzak2004,Labeur2004,Stastna2011}.


We present here, analytically supported by direct simulation,  that the spatial evolution of internal waves energy and the interplay between modes  over a patch of seabed undulations can be strongly dependent upon the distance of the patch to the neighboring seabed features. We show that accumulation of internal waves energy may be an order of magnitude larger at specific areas of a patch, solely based on \textit{where} the neighboring features are. The physics behind this phenomenon lies in the constructive and destructive interference of multiply reflected waves: If a patch of seabed undulations satisfies Bragg condition with internal waves, as mentioned above, it reflects part of the incident wave energy, but allows the rest to transmit. The transmitted wave then gets reflected back by the downstream reflector. But this reflected wave again reflects back by the patch of undulations via Bragg mechanism. This sequence of reflections continues indefinitely as multiply reflected waves add up and via constructive and/or destructive interference result in a very much different spatial distribution of energy over the patch than what is expected in the absence of the downstream topography. This phenomenon is a close cousin of the Fabry-Perot interference in optics through which two partially reflecting mirrors trap light \cite[][]{Fabry1897}. It has also been worked out in the context of surface gravity waves in a homogeneous (unstratified) fluid where many features similar to optics counterpart are found \cite[][]{couston2015,Yu2000}. In the context of internal waves in a continuously stratified fluid, nevertheless, the problem is significantly different as here Bragg resonance leads to the generation of an infinite number of internal wave modes simultaneously exchanging energy with each other through the seabed, creating a complex pool of interacting waves.

Real seabed topography in the ocean is usually composed of many Fourier components and, likewise, internal waves often arrive in a group forming a spectrum of frequency and wavenumber. Therefore, several interaction conditions may be satisfied simultaneously resulting in a substantial energy exchange that may lead to significant change in the spectral density function of internal waves. The sensitivity mechanism elucidated here sheds light on the importance of the details of topographic features on the resulting spatial distribution of wave activity, and may help pinpoint areas of the ocean where appreciable mixing is expected.

\section{Governing equations}\label{sec:goveq}
\begin{figure}
\centering
\includegraphics[height=1.2in]{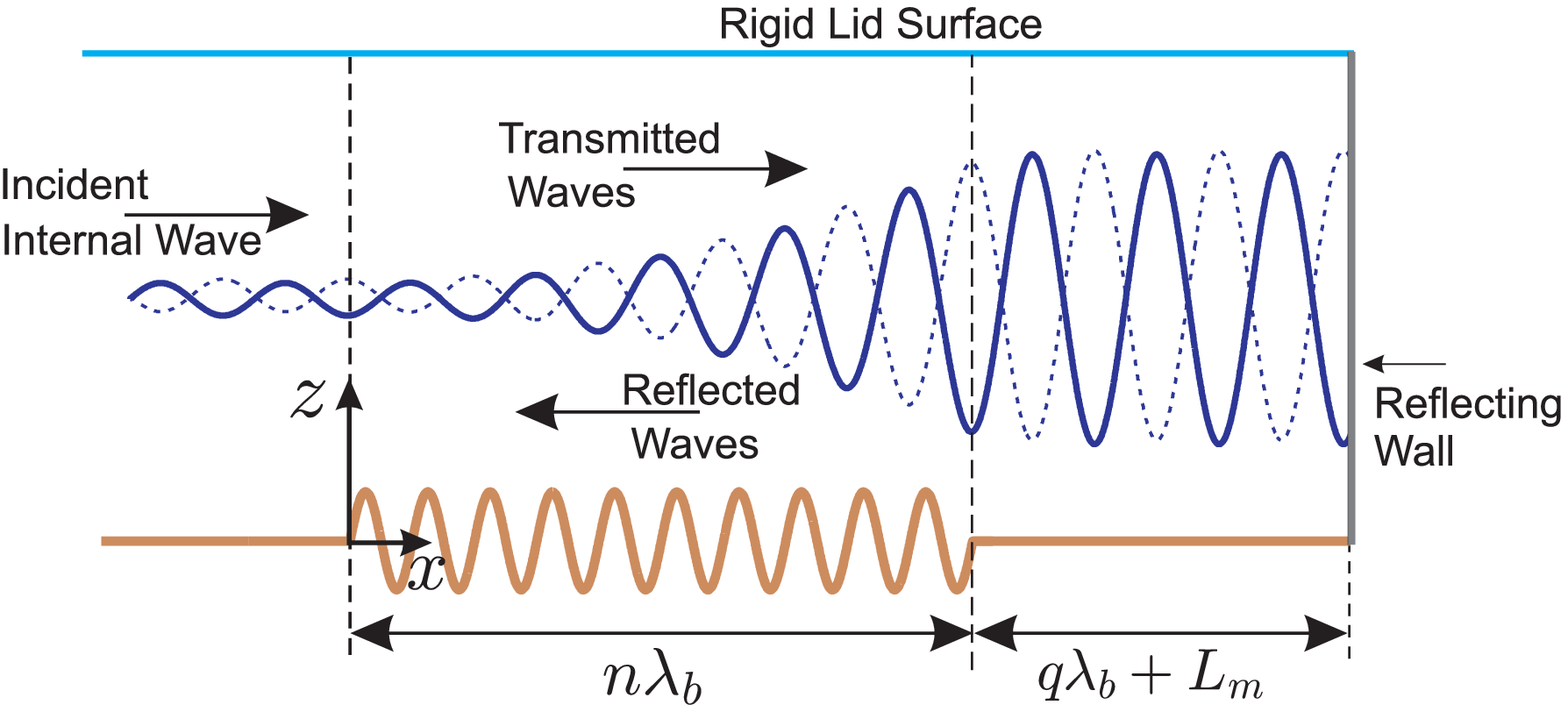}
\includegraphics[height=1.2in]{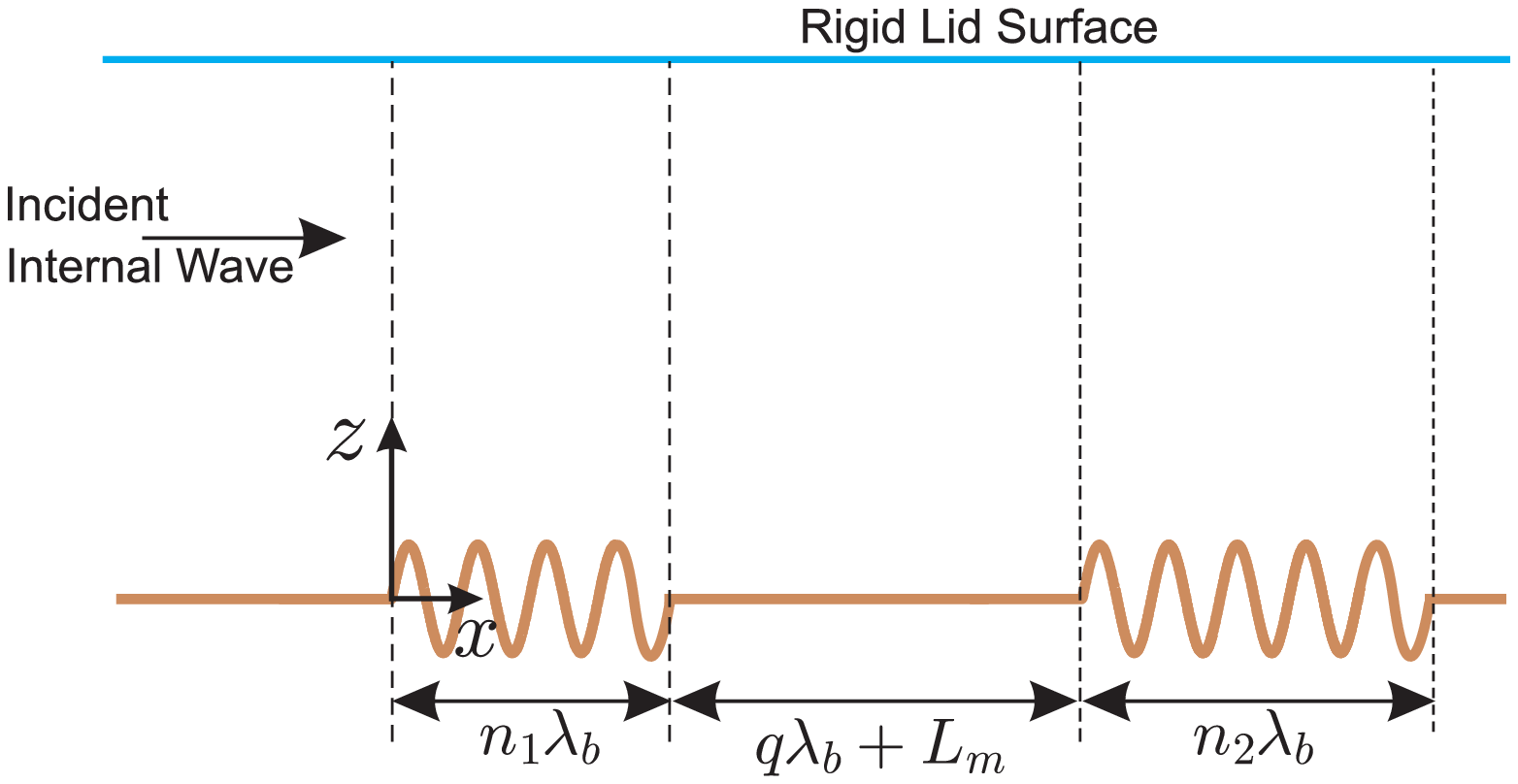}
	\put(-340,85){(a)} \put(-150,85){(b)}\\
\caption{Schematic representations of configurations considered here (a) an incident internal wave of wavelength $\lambda_i$ arrives from the far left to a patch of $n$ seabed ripples of wavelength $\lambda_b$. There is a reflecting wall at the distance $q\lambda_b+L_m$, ($L_m<\lambda_b$) measured from the end of the last ripple. (b) an incident internal wave of wavelength $\lambda_i$ arrives from the far left to two patches of $n_1$ and $n_2$ seabed ripples of wavelength $\lambda_b$ which are $q\lambda_b+L_m$ ($L_m<\lambda_b$) apart. We will show that the energy distribution over the patch and in the area between the patch and the wall (figure a), or between the two patches (figure b) strongly depends on $L_m$.}\label{fig:schematic}
\end{figure}

We consider an inviscid, incompressible, non-rotating, two-dimensional, and stably stratified fluid with small amplitude waves such that non-linear advection terms can be neglected. We put the Cartesian coordinate system on the seabed, with $z$ axis pointing upward (figure \ref{fig:schematic}). Density of this stably stratified fluid is $\rho(x,z,t)=\overline{\rho}(z)+\rho'(x,z,t)$, where $\overline{\rho}$ is the background density (density at equilibrium) and $\rho'$ is the density perturbation. Similarly, pressure is $p=\overline{p}(z)+p'(x,z,t)$ where $\overline{p} (z)$ satisfies the hydrostatic balance with the quiescent density as $\partial \overline{p}(z)/\partial z=-\overline{\rho}(z)g$, and $p'$ is the pressure perturbation. The governing equations for the velocity ${\bf u}=(u,w)$,  density and pressure perturbations $\rho', p'$ are \cite[e.g.][]{KUNDU[2012]} 
\bsa{gov}
   &\frac{\partial u}{\partial t}+\frac{1}{\rho_0}\frac{\partial p'}{\partial x} =0, \label{gova}\\
   &\frac{\partial w}{\partial t} +\frac{1}{\rho_0}\frac{\partial p'}{\partial z}= -\frac{\rho'g}{\rho_0}, \label{govb}\\
   &\frac{\p u}{\p x}+\frac{\p w}{\p z}=0,\label{govcc}\\
   &\frac{\partial \rho'}{\partial t}-\frac{N^2 \rho_0}{g}w = 0, \label{govc}
\esa
where $N=\sqrt {(-g/\rho_0) \partial \overline{\rho}/\partial z}$ is the buoyancy frequency and $\rho_0= \overline{\rho}(z=H)$ is the density at the free-surface. In \eqref{gov}, equations \eqref{gova} and \eqref{govb} are momentum equations, \eqref{govcc} is the continuity equation, and \eqref{govc} is obtained from conservation of energy. Assuming a rigid-lid condition at the surface $z=H$ and that the deviation of the seabed from the mean depth is given by $h(x)$, boundary conditions for the governing equations \eqref{gov} are 
\bsa{bc}
  	 w  = 0, ~~~ z=H; ~~~~ w = u  \frac{\partial h(x)}{\partial x}, ~~~ z=h (x).
\esa

Since the flow field is two-dimensional and divergence-free, the velocity can be written in terms of a stream function $\Uppsi (x, z , t)$ where $ u=\partial \Uppsi / \partial z$, and $w =-\partial \Uppsi/ \partial x$. Recasting governing equations \eqref{gov} in terms of $\Uppsi$, we obtain
 \ba\label{stre_wave}
\frac{ \partial^2}{\partial t^2} \left( \frac{\partial^2}{\partial x^2}  \Uppsi + \frac{\partial^2}{\partial z^2} \Uppsi\right) + N^2 \frac{\partial^2}{\partial x^2}\Uppsi=0, \,
 \ea
with the boundary conditions
  \ba\label{stre_wave_bc}
  \Uppsi (x, H, t) = \Uppsi (x, h(x), t)=0. \,
  \ea

We now consider time-harmonic solutions to  \eqref{stre_wave} in the form $\Uppsi(x, z, t)= \Re \lb \psi(x, z) e^{-\mathrm{i} \omega t} \rb $ where $\omega$ is the frequency of the motion. Considering a constant $N$, we define scaled horizontal and vertical variables $x^*=\mu\pi x/H$ and $z^*=\pi z/H$, where $\mu=\sqrt{ {\omega^2}/{(N^2-\omega^2)}}$. Using these scaled variables, the governing equation \eqref{stre_wave}, dropping asterisks, turns into
  \cite[e.g.][]{BUEHLER[2011]}
 \ba\label{stre_wave2}
 \frac{\partial^2}{\p x^2}  \psi  - \frac{\p^2}{ \partial z^2}\psi=0,
 \ea
Note that physical parameters (e.g. $N,H$) are hidden in the scaled variables. and do no appear explicitly. 
 

\section{Perturbation analysis} \label{sec:pert_an}
We use multiple scale perturbation analysis to solve for the wave field over a patch of small-amplitude ripples (i.e. $h(x)/H \ll 1$) in the area $0\leq x\leq L$. We assume that at the steady state wave field variables are functions of spatial variables $x,z$ and a slow horizontal scale $X=\epsilon x$ in which  $\epsilon\ll1$ is a  measure of the waves steepness.  We also assume that the solution to \eqref{stre_wave2}, $\psi$, can be expressed in terms of a convergent series, that is, 
\begin{eqnarray}\label{stream_expan}
\psi(x,z,X)=  \psi^{(0)} \left (x,z, X \right) + \epsilon \psi^{(1)} \left (x,z, X \right) + \mathcal{O} (\epsilon ^2). \,
\end{eqnarray}
Substituting \eqref{stream_expan} in equation (\ref{stre_wave2}) and collecting terms of the same order, at orders $\O(1)$ and $\O(\ep)$ we obtain, 
\bsa{920}
   &\mathcal{O} (1): \qquad \frac{\partial^2}{\p x^2} \psi^{(0)}  - \frac{\partial^2}{\p z^2} \psi^{(0)}=0, \, \label{stre_wave_ord_1}  \\
   & \mathcal{O} (\epsilon): \qquad \frac{\partial^2}{\p x^2} \psi^{(1)}  - \frac{\partial^2}{\p z^2}\psi^{(1)}=-2\epsilon \frac{\partial^2}{\p x \p X}\psi^{(0)}. \, \label{stre_wave_ord_eps}
\esa

In a search for wave solutions to the original equation \eqref{stre_wave}, we consider the following general solution to equation \eqref{stre_wave_ord_1}
\ba\label{order_zero_sol}
\psi^{(0)} \left (x,z, X \right) = \sum_{m=1}^{\infty} \left (\widehat{T}_m (X) e^{\i mx} + \widehat{R}_m (X) e^{-\i mx}\right) \sin m z , \,
\ea
where $\widehat{T}_m, \widehat{R}_m$ are amplitudes of transmitted and reflected waves respectively. The specific form of solution \eqref{order_zero_sol} assumes that these amplitudes can slowly vary over the patch of seabed corrugations. Upon substitution of \eqref{order_zero_sol} in (\ref{stre_wave_ord_eps}) we obtain
\begin{align}\label{stre_wave_ord_eps1}
\frac{\p^2\psi^{(1)}}{\p x^2}-\frac{\p^2 \psi^{(1)}}{\p z^2}=-2\i\sum_{m=1} ^{\infty}m\left(\frac {\p \widehat{T}_{m}(X)}{\p X} e^{\i mx}-\frac {\p \widehat{R}_{m}(X)}{\p X} e^{-\i mx} \right) \sin mz.
\end{align}
Coefficient $\p {\widehat{T}_{n}}/\p X$ ( $\p {\widehat{R}_{n}}/{\p X}$) is readily obtained by multiplying both sides of \eqref{stre_wave_ord_eps1} by $e^{-\i nx} \sin nz$ ( $e^{\i nx} \sin nz$) and integrating over $x \in [-\pi, \pi]$ and $z\in [0,\pi]$:
 \ba\label{stre_wave_ord_eps4}
 \frac{\p \widehat{T}_{n}}{\p X}=\frac{-\i}{2 \pi^2} \int_{x=-\pi} ^{\pi} \psi^{(1)} (x, 0,X) e^{-\i nx} \d x ,~~   \frac{\p \widehat{R}_{n}}{\p X}=\frac{\i}{2 \pi^2} \int_{x=-\pi} ^{\pi} \psi^{(1)} (x, 0,X) e^{\i nx} \d x , 
 \ea
where integration by parts is used for the left hand side of \eqref{stre_wave_ord_eps1}. Taylor expansion of the boundary condition \eqref{stre_wave_bc} at $z=0$ yields $\psi^{(1)} (x, 0, X) = -h(x) \psi_z^{(0)} (x,0, X)$, and therefore
 \begin{align}
   \frac{\p}{\p X}\left\{ \begin{array}{c}
       \widehat{T}_{n} \\
       \widehat{R}_{n}
   \end{array}\right\}
 =\frac{\pm \i}{2 \pi^2} \int_{x=-\pi} ^{\pi} h(x) \sum_{m=1} ^{\infty}m\left(\widehat{T}_m(X) e^{\i mx}+\widehat{R}_m(X) e^{-\i mx} \right)  e^{\mp \i nx} \d x ,   \label{TR_stre} 
\end{align}
where the upper/lower signs are respectively for $\widehat{T}_{n}$ and  $\widehat{R}_{n}$.
We can write $h(x) =\sum_{j= -\infty}^{\infty} h_j e^{\i jx},\ $ therefore,
\begin{align}\label{stre_coef_sol2}
   \frac{\p}{\p X} \left\{ \begin{array}{c}
       \widehat{T}_{n} \\
       \widehat{R}_{n}
   \end{array}\right\} &=\frac{\pm \i}{2 \pi^2} \sum_{j=-\infty}^{\infty}\sum_{m=1}^{\infty} \int_{x=-\pi}^{\pi}  m h_j  \left(\widehat{T}_m(X)  e^{\i(j\mp n+m)x}  +\widehat{R}_m(X) e^{\i(j\mp n-m)x} \right) \d x  \nn \\
   &=\frac{\pm \i}{\pi} \sum_{j=1}^{\infty}\sum_{m=1}^{\infty}   m h_j  \left(\widehat{T}_m(X) \delta_{j,\pm n-m} +\widehat{R}_m(X)  \delta_{j,\pm n+m} \right)  \nn \\
   &=\frac{\pm \i}{\pi}\sum_{m=1}^{\infty} m\left(h_{\pm n-m}\widehat{T}_m(X) +h_{\pm n+m}\widehat{R}_m(X)  \right).   
\end{align}
The velocity transmission and reflection coefficients ($T_m, R_m$) are obtained from $\widehat{T}_m, \widehat{R}_m$ through $T_m=-\i m\widehat{T}_m (X)$ and  $R_m=\i m\widehat{R}_m (X)$. The spatially averaged kinetic and potential energy for each mode are obtained from $ \left<E^k_n \right>=(1/2)\rho_0(1/\lambda_n) \int_{0}^{\lambda_n} \mathrm{d}x \int_{0}^{H} (\overline{u_n^2+w_n^2})~ \d z$ and $\left<E^p_n \right>= (1/\lambda_n)\int_{0}^{\lambda_n} \mathrm{d}x \int_{0}^{H} (g^2 \overline{\rho_n'^2})/(2\rho_0 N^2) ~\d z$, where  the overbar  denotes the temporal average and $\left< ~.~ \right>$ shows the spatial average. These equations result in
\begin{eqnarray}
\left<E^k_n \right>=\left<E^p_n \right>= \frac{1}{8} \rho_0 A_n^2
 \frac{N^2}{\omega^2} H = \frac{1}{8} \rho_0 \left(T^2_n+R^2_n \right) \left(1+\frac{m_n^2}{k_n^2} \right) H,\,
\end{eqnarray}
where $m_n,k_n$ are vertical and horizontal wavenumbers respectively, $\lambda_n=2\pi/k_n$. Hence, the total energy per unit area is 
\begin{eqnarray}
\left<E \right>=\left <E^k \right>+\left<E^p \right>= \sum_{n=1}^{\infty}\frac{1}{4} \rho_0 \left(T^2_n + R^2_n \right) \left(1+\frac{m_n^2}{k_n^2} \right) H.  
\end{eqnarray}
%
We define normalized total energy by using the mode one incident internal wave energy as the reference, i.e.,
\begin{align}\label{energy_length}
\widetilde{E} =\frac{\left<E \right>}{\left<E_{incident} \right>} =\sum_{n=1}^{\infty} \frac{T_n^2 + R_n^2}{T_{1,(x=0)}^2}. 
\end{align}

\section{Results and Discussion} \label{sec:prob_set}
With the formulation of \S3 in hand, we now proceed to study the spatial evolution of the internal wave energy over a patch of seabed ripples. For the sake of completeness, we review the energy distribution over a single patch of ripples, and then focus our attention on: \textit{i.} a patch of seabed ripples located at distance $q\lambda_b+L_m$ from a vertical wall, and \textit{ii.} two patches of ripples at the distance $q\lambda_b+L_m$ from each other ($q$ being an integer number). We show that in both cases,  amplitudes of different mode internal waves and the overall energy distribution strongly depend on $L_m$.

\subsection{Single patch} \label{sec:sing_patch}

In a continuously stratified fluid of constant $N$, and if normalization of \S2 is employed, then a frequency $\omega$ is associated with an infinite number of internal wave modes with integer wavenumbers. If an internal wave mode $m$ propagates over a seabed undulation that has a component with the wavenumber $n_b=2m$, then through Bragg resonance new free-propagating internal waves of mode $m\pm n_b$ are excited (resonated). These two new waves can interact with the same topography to generate yet a new set of resonant waves $m+2n_b$, and $m-2n_b$. Eventually, and if the patch is long enough, an infinite number of waves with wavenumbers $m\pm jn_b$, with integer $j\in (0,\infty)$, will appear in the water.

\begin{figure}
 	\centering
 		{\includegraphics[width=3in]{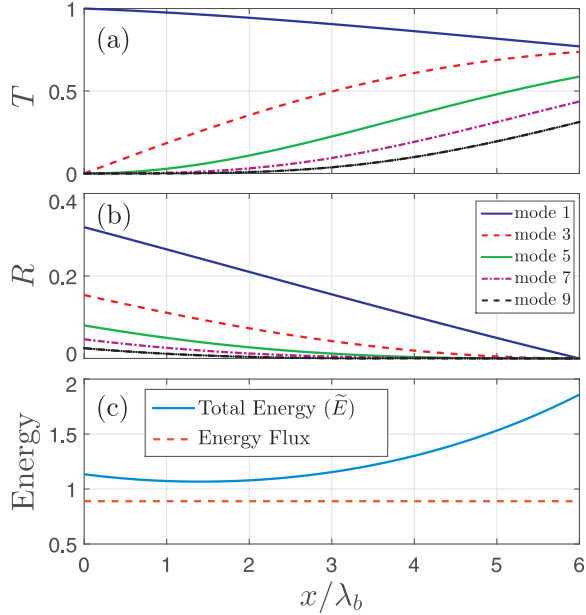}}
 		\caption{Interaction of a mode one incident internal wave ($m=1$) with a single patch of sinusoidal ripples $h(x)=0.04 \pi \sin(2x)$ ($0\le x\le 6\lambda_b$, i.e. the patch is composed of six ripples). Figures a, b and c respectively show transmission coefficient $T$, reflection coefficient $R$ and the normalized energy per unit area $\widetilde E$ over the patch. Energy of the incident wave (mode 1) decreases as energy goes to higher modes in transmission, as well as the mode one and higher modes in reflection. The overall energy per unit area $\widetilde E$ initially decreases a little, but eventually takes off toward the downstream of the patch. Energy flux (dashed line in figure c) is constant over the patch, as expected.} \label{fig:six_ripples}
 \end{figure}

For illustration purposes, let us consider a mode one (i.e. $m=1$) internal wave propagating over a monochromatic patch of ripples $h(x)=a_b \sin n_b x$, with $a_b=4\pi/100$ (which implies the ripples amplitude is 4\% of the water depth) and $n_b=2$. We consider a patch that extends over the area  $0\le x\le L=6 \lambda_b$ where $\lambda_b=2\pi/n_b$ is the seabed ripples' wavelength. The variation of the amplitude of the first four resonated waves along with the amplitude of the incident wave is shown in figure \ref{fig:six_ripples}. Incident wave of mode $m=1$ arrives from $-\infty$, and upon interaction with the seabed $n_b=2$, generates new waves with wavenumbers $m+n_b=3$ and $m-n_b=-1$ (the negative sign shows that this new wave, which is mode 1, moves in the opposite direction and hence appears in the reflection plot, i.e. figure \ref{fig:six_ripples}b). These newly generated waves pick up amplitude at the cost of incident wave amplitude decaying over the patch, as is seen in figure \ref{fig:six_ripples}a. Once the amplitude of mode 3 wave (red dashed line) is large enough, through the same topography, mode 5 is resonated, and the interaction goes on. Similar story holds for the waves in reflection. Mode one wave in reflection resonates mode 3 and so on. While equation \eqref{stre_coef_sol2} gives us all modes that are generated here, we only presented the first four wave modes (plus the incident). Figure \ref{fig:six_ripples}c shows the energy per unit area in the water column. Since group velocity of higher modes is slower, energy is \textit{accumulated} toward the end of the patch where more energy is in higher modes that travel \textit{slower}. As expected, in the steady state energy flux remains unchanged (energy flux is normalized by the energy flux of incident wave). Note that energy density per unit area everywhere is greater than the incident wave energy density per unit area, and toward the end of the patch becomes much higher. This is clearly a result of generation of internal waves with higher wavenumbers. 

\subsection{Patch-wall case} \label{sec:wall}

Now let us assume that there is a wall on the downstream of the patch, at the distance $q\lambda_b+L_m$ from the end of the last ripple (c.f. figure \ref{fig:schematic}a). As waves propagate over the patch, a picture similar to figure \ref{fig:six_ripples} starts to form. Waves on the downstream, nevertheless, are reflected back by the wall and start to interact again with the topography. These left propagating waves are partially transmitted, but also partially reflected back toward the wall. It turns out that the resulting effect is very complicated and a strong function of $L_m$. 

\begin{figure}
	\centering
	\includegraphics[width=5in]{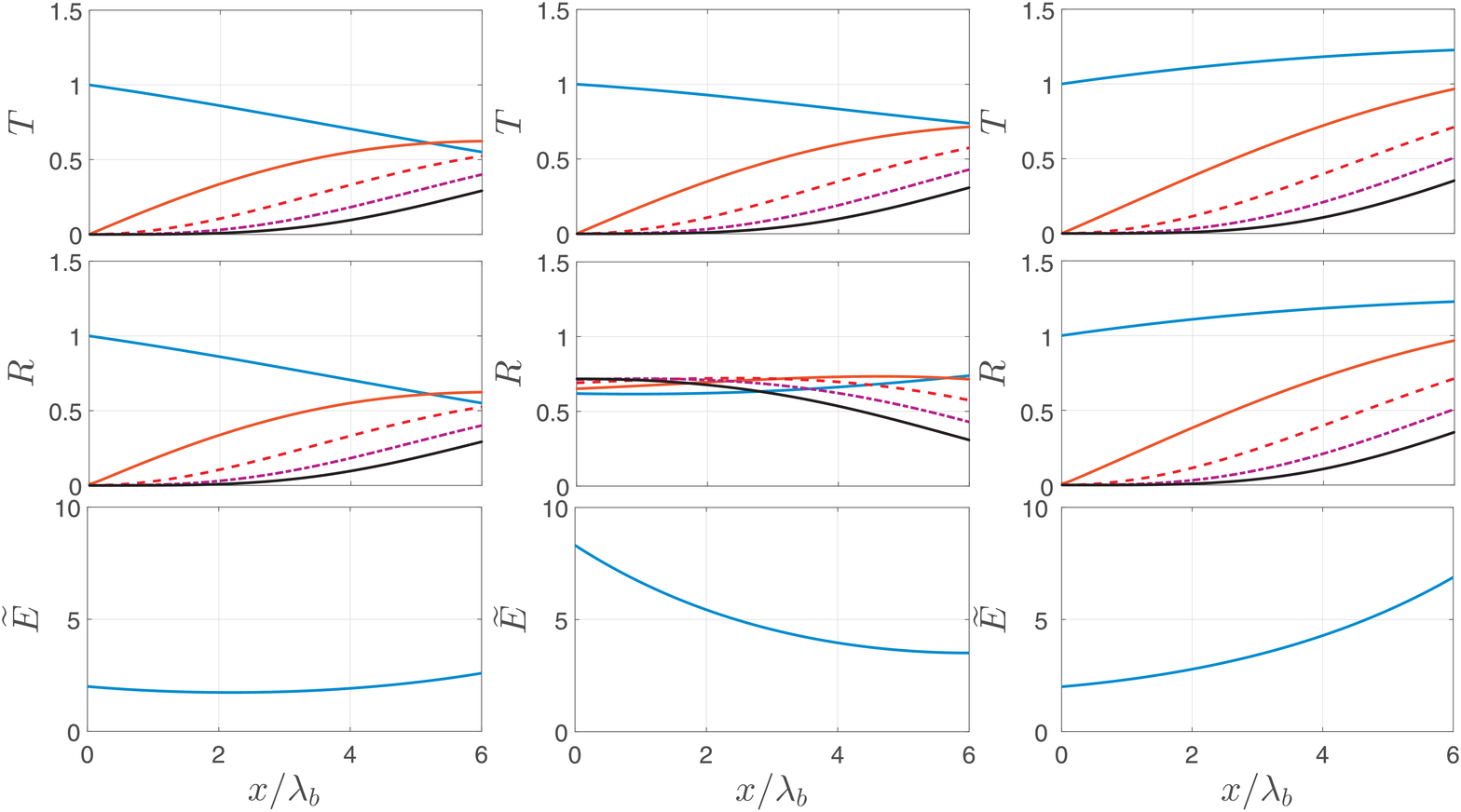}
	\put(-295,190){(a)} \put(-178,190){(b)} \put(-58,190){(c)}\\
	\caption{Variation of transmission coefficient ($T$), reflection coefficient ($R$) and the normalized energy per unit area $\widetilde E$ over the patch of $n$=6 ripples, for a downstream wall at the distance (a) $L_m/\lambda_b$=0, (b) $L_m/\lambda_b$=0.25, and (c) $L_m/\lambda_b$=0.50, from the end of the patch.  Plotted are mode 1 internal wave (\blue{\L}), mode 3 (\red{\L}), mode 5 (\red{\dashL}), mode 7 (\magenta{\cdotL}), and mode 9 (\black{\L}). Higher modes exist, but are not shown here. Note that for $L_m/\lambda_b$=0, 0.5 (figures a , c) through a complicated set of chain interactions all the energy eventually goes back to mode 1 on the upstream side of the patch. In this case an upstream observer does not see any trace from the patch of ripples. To this observer, everything looks like a perfect reflection from the wall in the absence of seabed irregularities. For any other value of $L_m/\lambda_b$, the upstream observer sees many other internal wave modes besides mode 1.}\label{fig:wall_distance_t_r}
\end{figure}

We present in figures \ref{fig:wall_distance_t_r}a-c the final steady state transmission/reflection amplitudes of different modes and energy per unit area over a patch of six ripples with a wall at the distance $L_m/\lambda_b$=0, 0.25 and 0.50 respectively. Other parameters of the ripples are the same as in \S4.1. For $L_m/\lambda_b$=0, energy goes from mode one to higher modes as the incident wave propagates over the patch. However, interestingly after reflection the energy entirely goes back to mode 1 such that in the upstream there is no reflected wave except mode 1. Energy per unit area $\widetilde E$ does not change much over the patch. The spatial evolution of modes for the case of $L_m/\lambda_b$=0.5 is similar to the case of $L_m/\lambda_b$=0, except that in the former the amplitude of mode 1 wave \textit{increases} over the patch, resulting in a significant energy increase over the patch toward the downstream side. For the distance $L_m/\lambda_b$=0.25, the transmitted figure is qualitatively similar to the case of $L_m/\lambda_b$=0, but the reflection figure is very much different: higher modes remain with non-zero amplitude (with finite energy) at the beginning of the patch and propagate upstream. This means that higher modes can be seen upstream of the patch moving toward left (this is not the case for $L_m/\lambda_b$=0, 0.5). In this case, $\widetilde E$ is highest at the beginning of the patch and decays fast toward the wall side of the patch. Note that the spatial distribution of energy is periodic with the wavelength $\lambda_b$ and this can be shown to be also the case for each of wave modes involved. Therefore addition of $q\lambda_b$ ($q$ being an integer number) to the distance between the patch and the wall does not affect the results shown here.

\begin{figure}
	\begin{center}
		\includegraphics[width=2.3in]{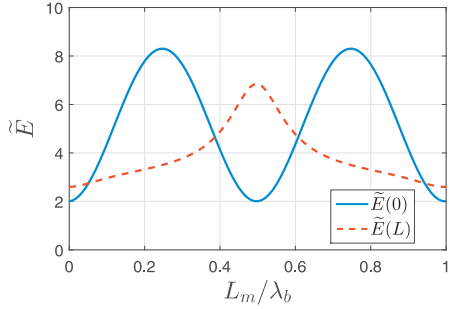}
		\caption{Spatial variation of energy per unit area $\widetilde E$ over the patch of bottom ripples ($n$=6) for different distance of the patch from a reflecting wall downstream.  The energy $\widetilde E$ is maximum at the beginning (upstream side) of the patch for $L_m/\lambda_b$=0.25  and 0.75, and is maximum at the wall side (downstream side) of the patch for $L_m/\lambda_b$=0.5.}\label{fig:wall_distance}
	\end{center}
\end{figure}

To see the behavior of energy density per unit area for various $L_m$, figure \ref{fig:wall_distance} shows energy at the beginning of the patch (solid blue line) and at the end of the patch (dashed red line) as a function of $L_m$. For $L_m/\lambda_b$=0, 0.5 we, in fact, obtain minimum energy at the beginning of the patch. As shown in figure \ref{fig:wall_distance_t_r}, in both cases only mode one wave appears upstream: incident and reflected waves that together form a mode one standing wave upstream of the patch. Energy density near the wall, however, is maximum for $L_m/\lambda_b$=0.5 and minimum for $L_m/\lambda_b$=0. The other important extremum is $L_m/\lambda_b$=0.25 for which $\widetilde E$ is maximum upstream as, in addition to mode one, several higher mode waves also reflect back toward the $-\infty$. The behavior of energy at upstream is symmetric about $L_m/\lambda_b$=0.5. Also seen in figure \ref{fig:wall_distance} that for $n$=6, $\widetilde E(0)$ may be affected by a factor of $\sim 4$ depending on $L_m$. For $n$=12, it turns out this contrast is as big as $~50$ times.

To provide an independent cross-validation to the obtained results, we present here direct simulation via SUNTANS (the Stanford Unstructured Nonhydrostatic Terrain-following Adaptive Navier-Stokes Simulator). SUNTANS is a finite-volume solver developed for simulation of three-dimensional non-hydrostatic internal waves in the ocean \cite[][]{FRINGER[2006]}. Since its introduction in 2006, SUNTANS has undergone cross-checks extensively  \cite[e.g.][]{FRINGER[2008],WANG[2009],KANG[2012]}.

\begin{figure}
  	\centering
  	\includegraphics[width=1.6in]{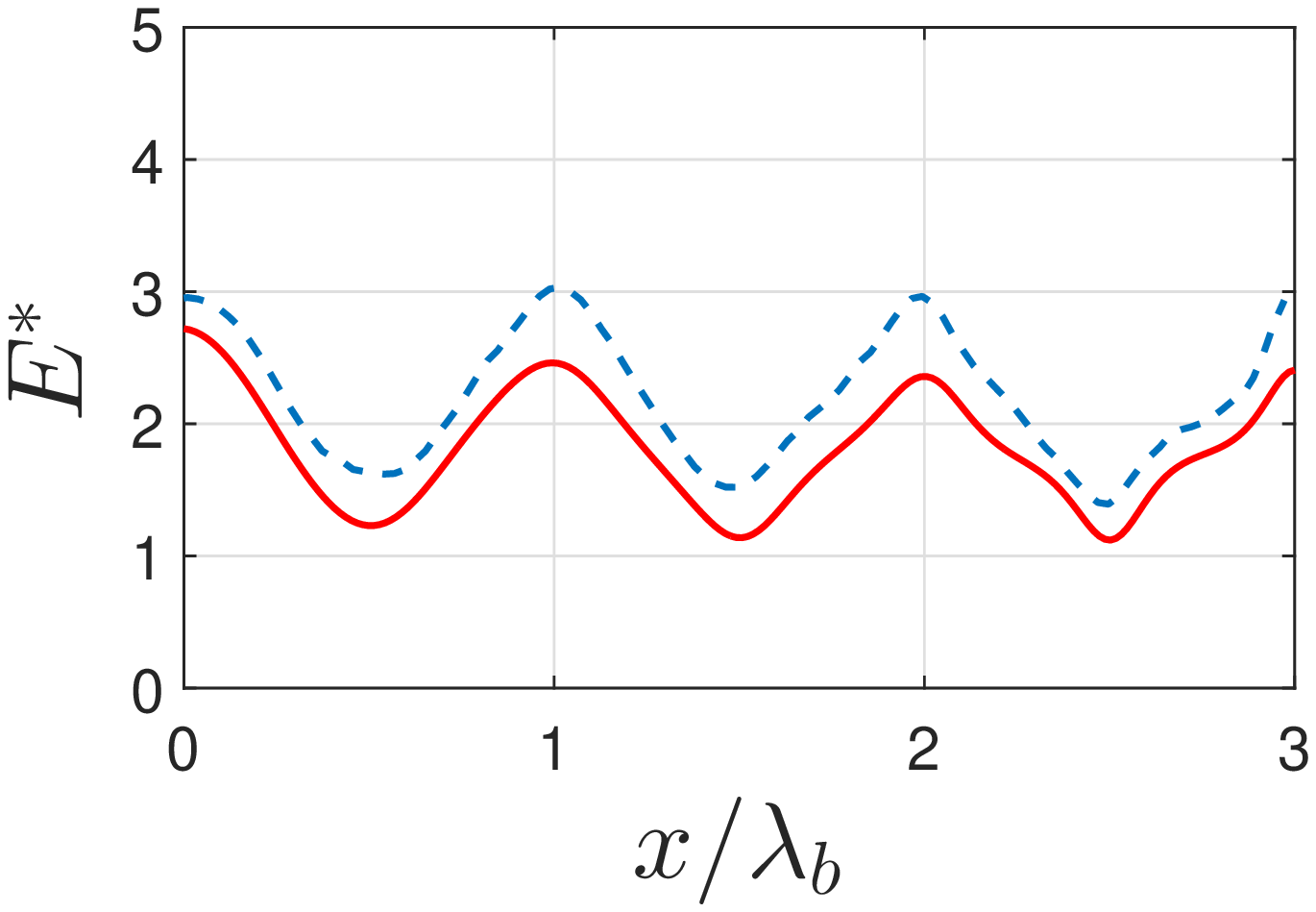}  	  \includegraphics[width=1.6in]{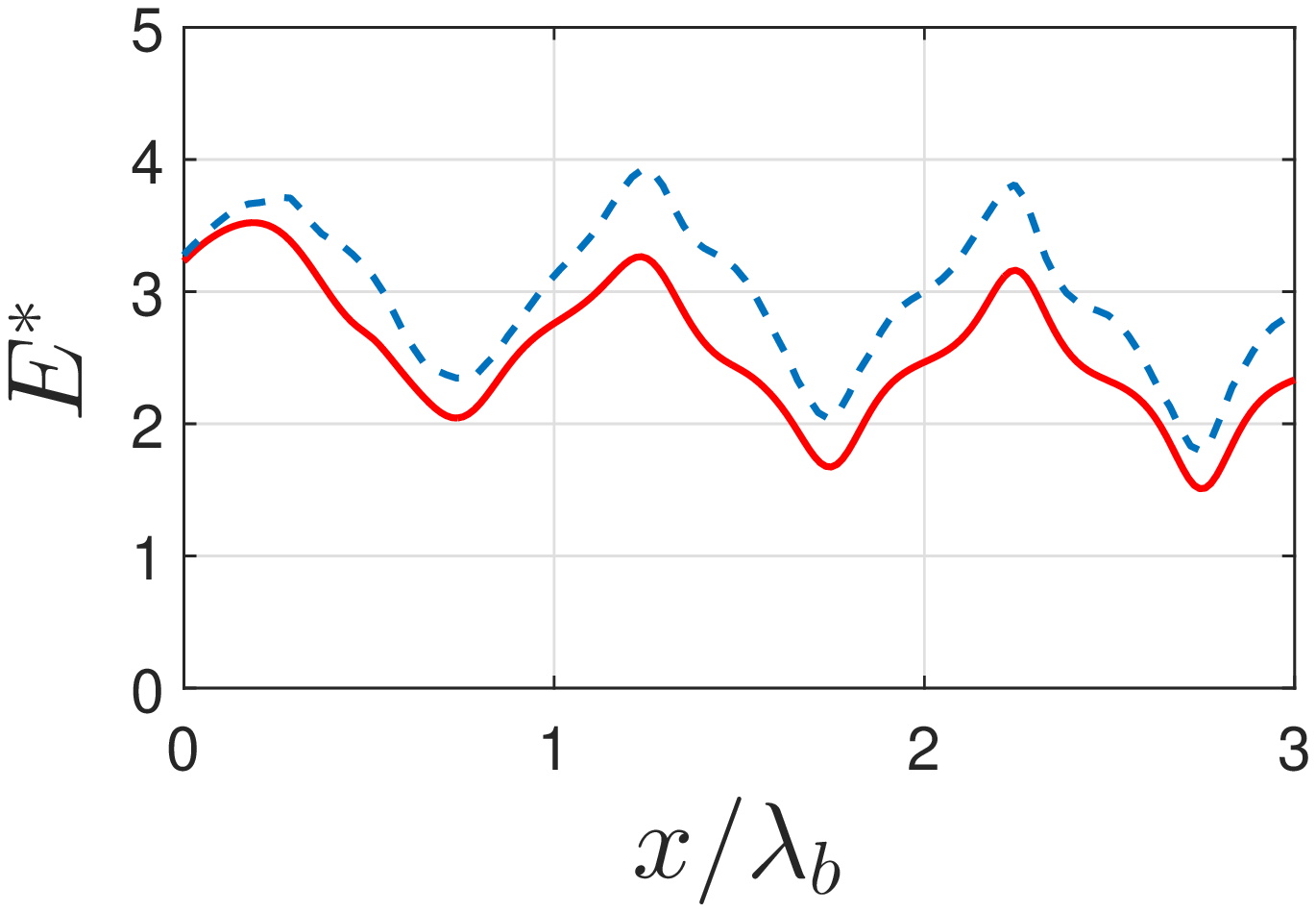}
  	\includegraphics[width=1.6in]{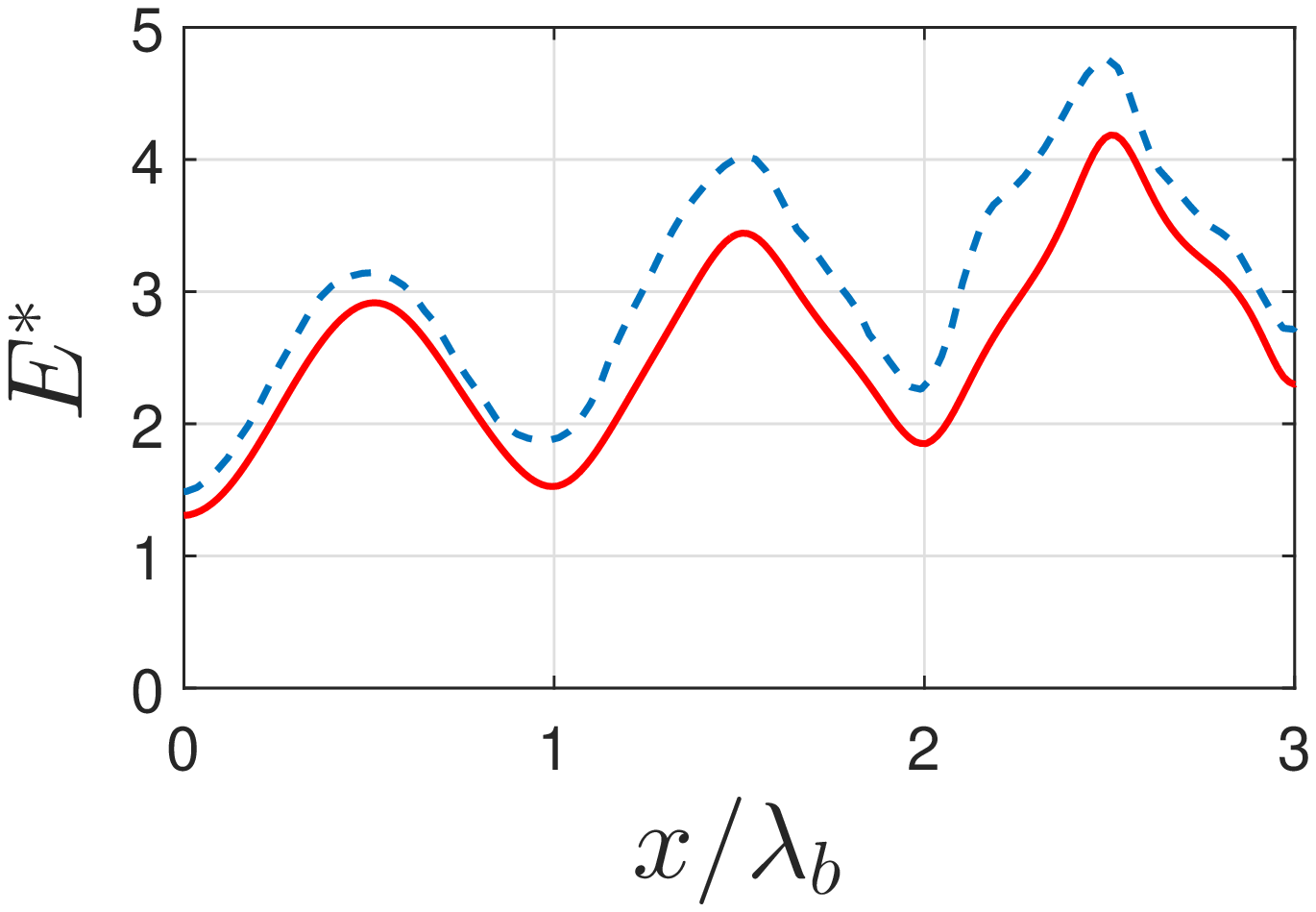}
  	\put(-335,62){(a)} \put(-215,62){(b)} \put(-95,62){(c)}\\
  	\caption{Comparison of energy per unit area ($E^*$) from analytical solution (\red{\L}), and  direct simulations (\blue{\dashL}) for respectively $L_m/\lambda_b$=0, 0.25 and 0.5 in figures a,b and c.}\label{fig:suntans}
  \end{figure}

We consider a two-dimensional domain with a constant Brunt-V\"ais\"al\"a frequency of $N = 0.0443 $ s\textsuperscript{-1} (i.e. 2\% change in density over the chosen depth of $H$=100 meters). At the left boundary, mode one internal wave is imposed through specifying horizontal velocity according to $U (0, z, t) = U_0 \cos (m_1 z) \cos (\omega t)$  where $\omega= 0.0266$ s\textsuperscript{-1}, $m_1=0.0315$ m\textsuperscript{-1}  is the first mode vertical wavenumber, and $U_0= 0.013$ m/s. Other boundary conditions are chosen as free-slip at the bottom, a solid wall with no-normal velocity at the right-end boundary, and a free surface on the top \cite[for a detailed discussion of the effect of free surface vs rigid-lid see][]{couston2016}. Domain length is $L = 20 \lambda_i = 5326.4$ m ($\lambda_i$ is the wavelength of the incident wave) and chosen such that there is enough time for the steady state to be reached. The grid resolution is $1000 \times 100$ in respectively $x$ and $z$ directions. A patch of three ripples  on the seabed with the amplitude $4$ m is considered. The comparison of spatial distribution of energy from theoretical predictions and those obtained by direct simulation via SUNTANS is shown in figure \ref{fig:suntans} for $L_m/\lambda_b$=0, 0.25 and 0.50. In this figure, $E^*$ is the total energy normalized by the total energy of the incident wave ($E_i$) calculated as $E^*=(E^k+E^p)/E_i$, where the kinetic energy ($E^k$) and the potential energy ($E^p$) are $ E^k =\sum_{n=1}^{\infty}1/2\rho_0  \int_{0}^{H} (\overline{u_n^2+w_n^2})~ \d z$ and $E^p = \sum_{n=1}^{\infty} \int_{0}^{H} (g^2 \overline{\rho'^2_n})/(2\rho_0 N^2) ~\d z$ (Note that $\widetilde E$ is the spatial average of $E^*)$. As can be seen, theoretical predictions and direct simulation results are in good agreement with each other. The small discrepancy is attributed to the effect of the free surface, that an initial value problem is solved by SUNTANS, and that theoretical energy is calculated from $\O(1)$ velocity.

\subsection{Two patches case} \label{sec:two_patch}

\begin{figure}
	\begin{center}
		\scalebox{0.44}{\includegraphics{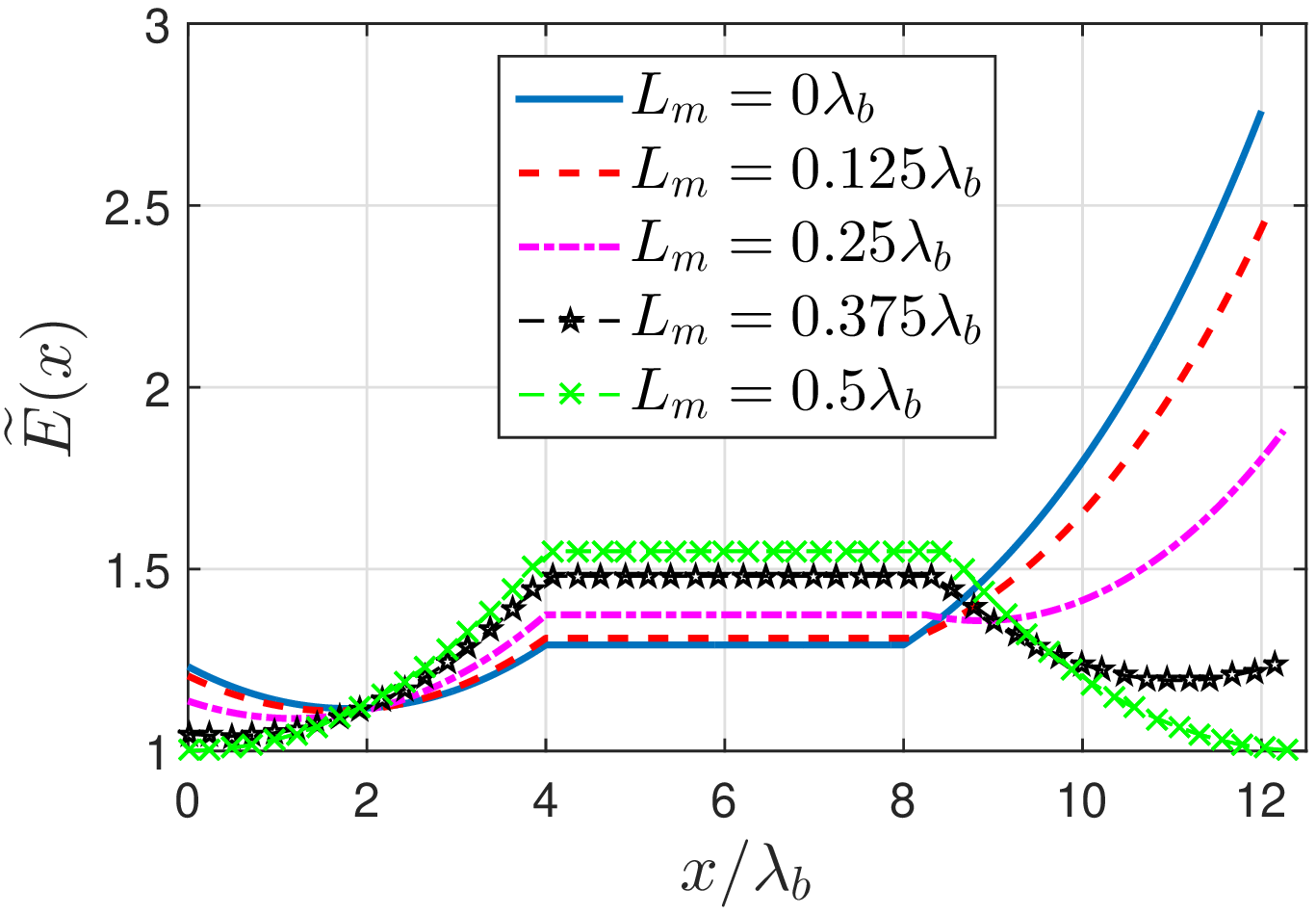}}
		\put(-160,105){(a)}
		\scalebox{0.36}{\includegraphics{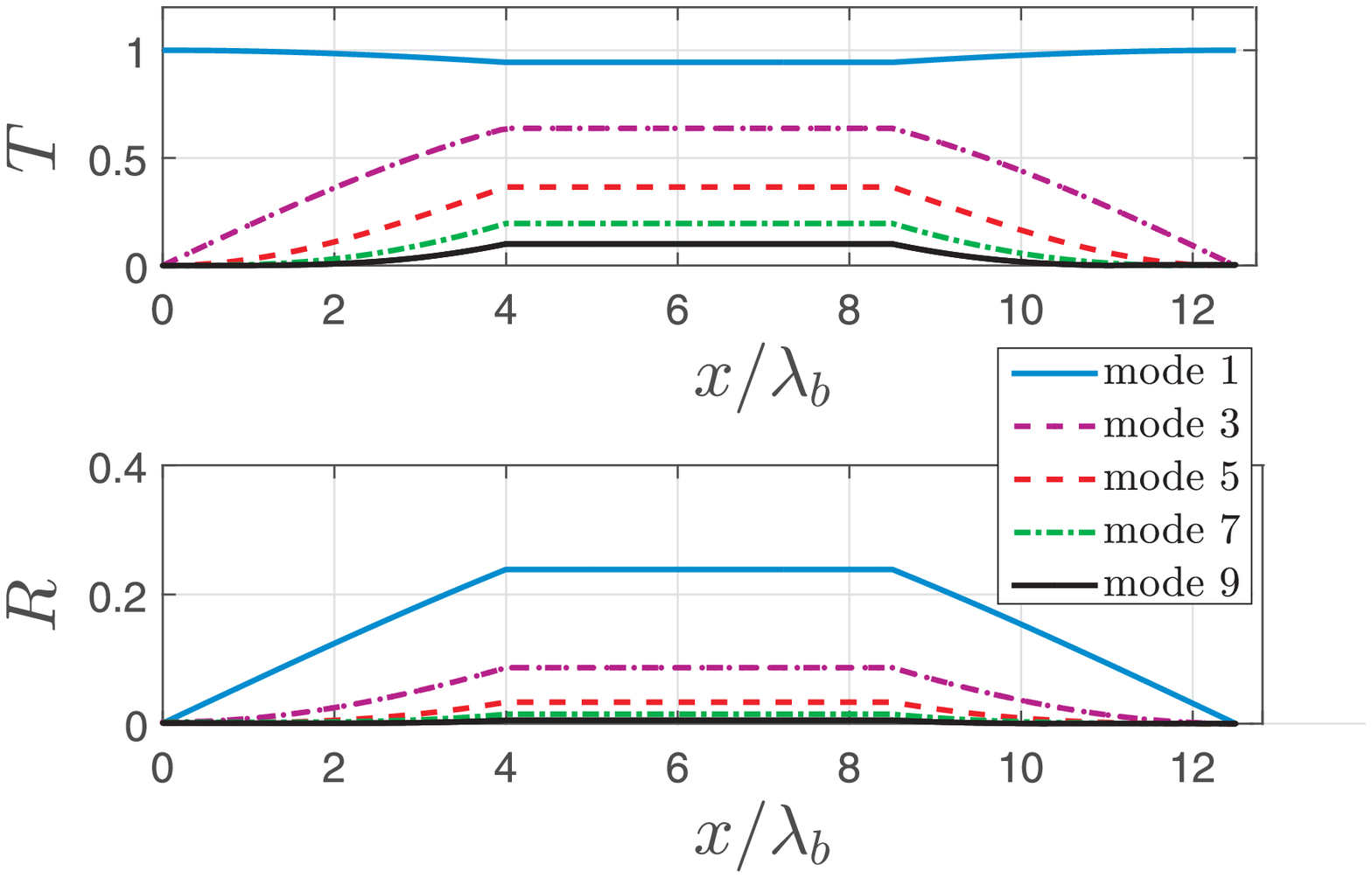}}
		\put(-180,110){(b)} \put(-180,59){(c)}\\ \vspace{-0.1cm}
		\includegraphics[width=1.73in]{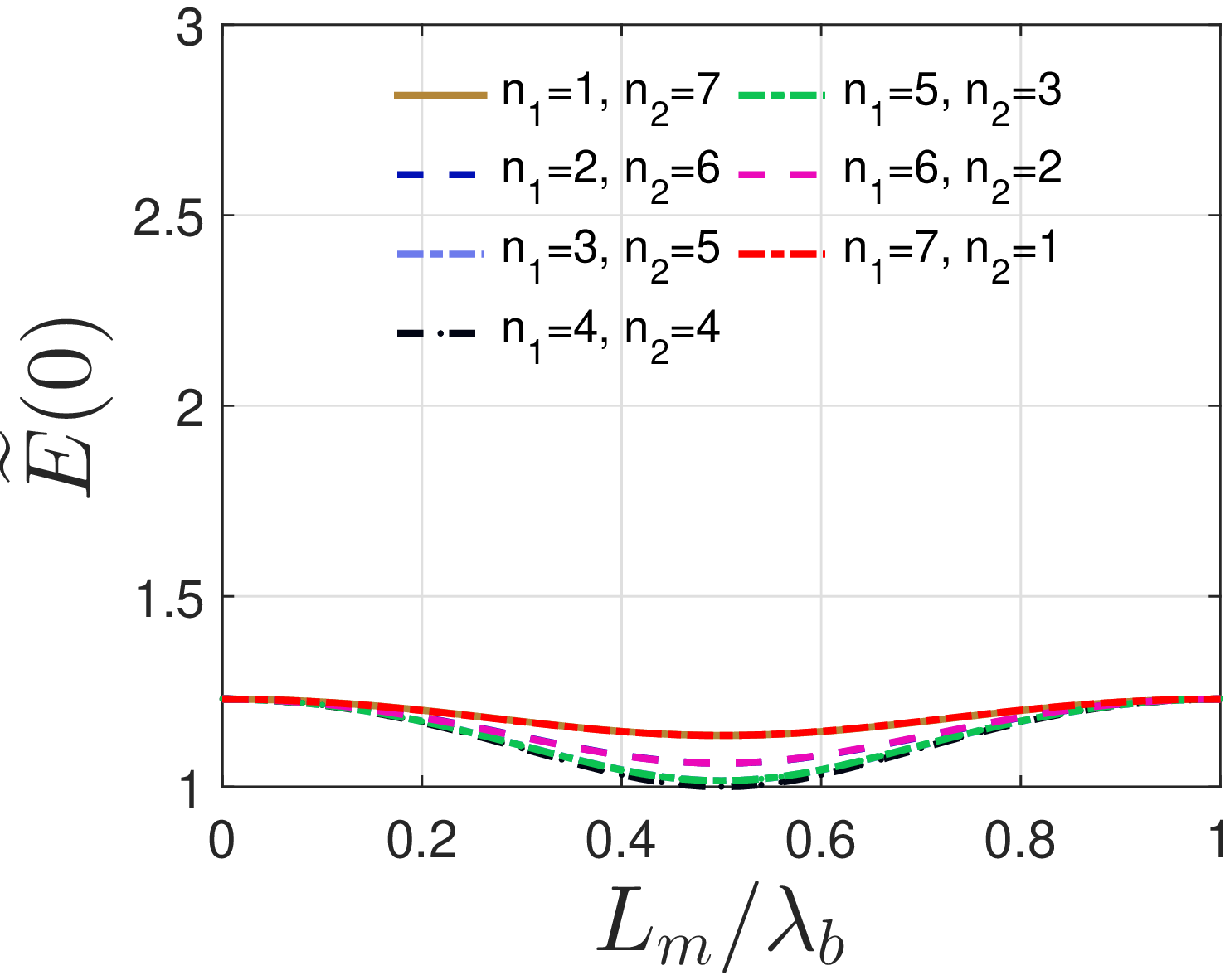}
		\includegraphics[width=1.73in]{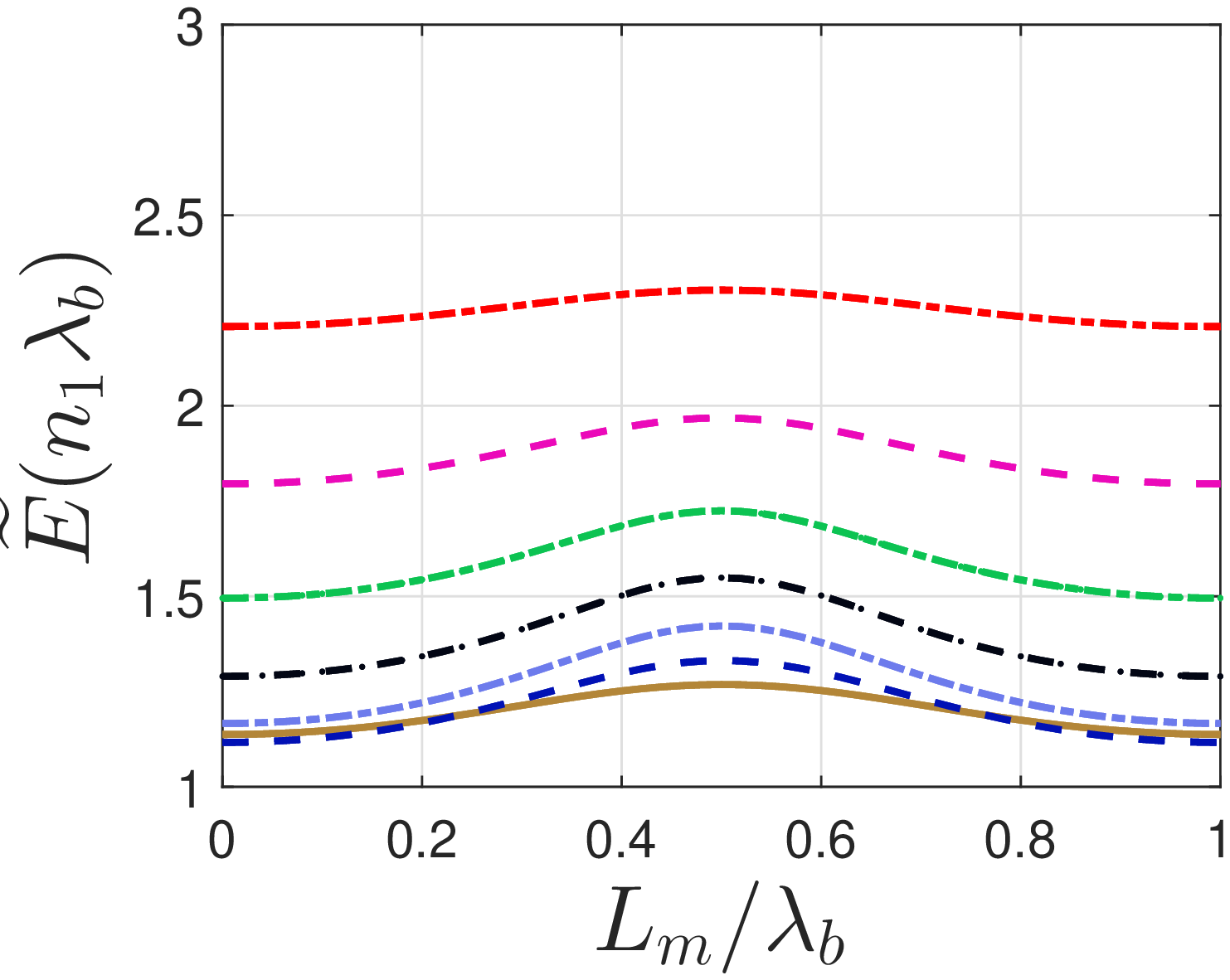}
		\includegraphics[width=1.73in]{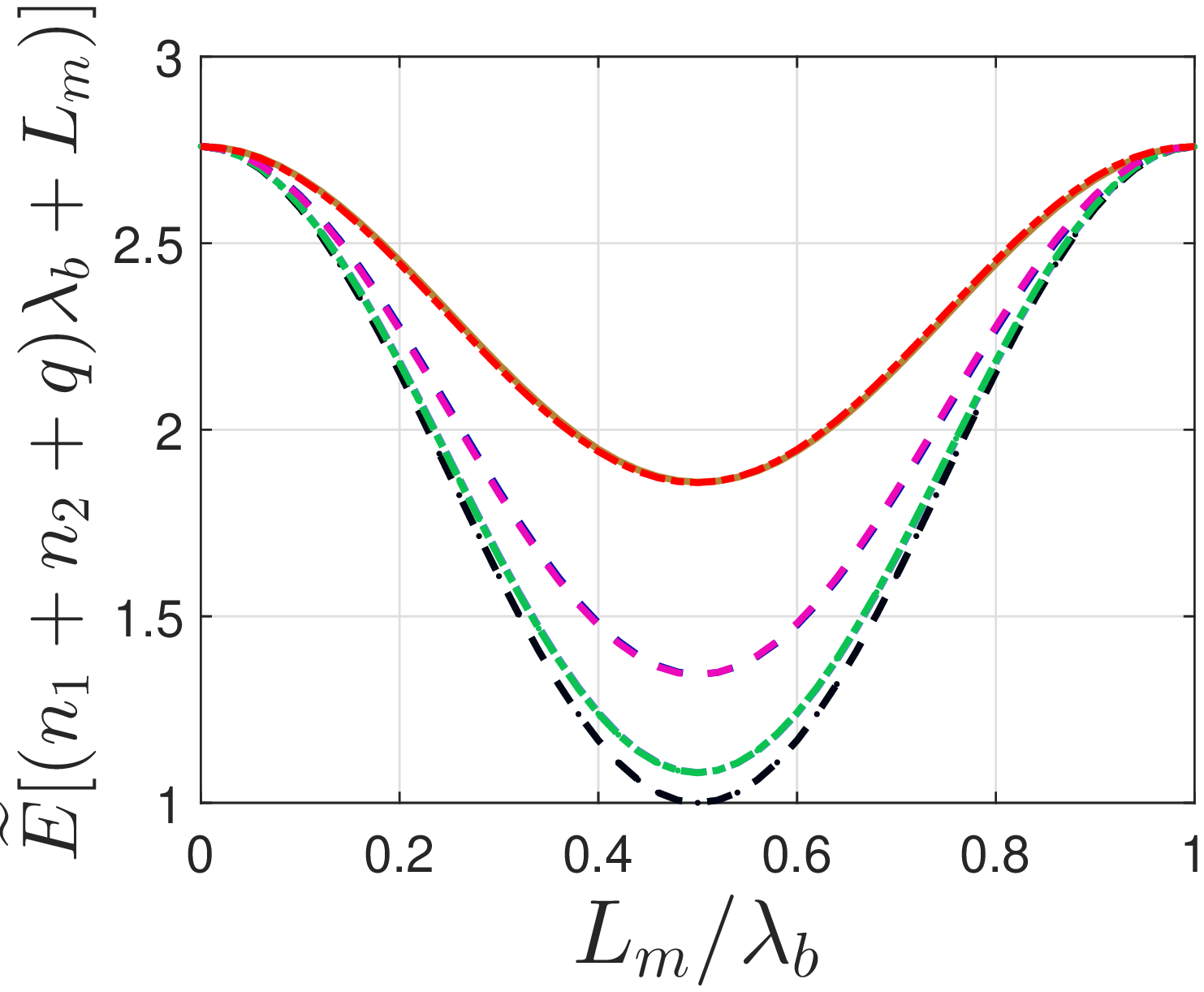}
		\put(-360,75){(d)} \put(-227,75){(e)} \put(-82,75){(f)}\\
		\caption{(a) The spatial evolution of energy density $\widetilde E$ in a two patch system with $n_1=n_2=4$ (c.f. figure 1) for different distances between the two patches $L_m/\lambda_b$= 0, 0.125, 0.25, 0.375, and 0.5. The normalized amplitude of seabed corrugations is 0.04. The maximum energy between the two patches is obtained when $L_m/\lambda_b=0.5$ and the maximum energy at the end of the second patch obtains for $L_m/\lambda_b$ =0. (b,c): Spatial evolution of amplitude of first five resonant modes (modes 1, 3, 5, 7 and 9) over the two patches of ripples with $L_m/\lambda_b=0.5$. (d,e,f): Evolution of energy as a function of number of ripples at (d) beginning of the first patch, (e) between two patches and (f) end of the second patch. }\label{fig:two_patch_ener}
	\end{center}
\end{figure}

Now let us consider a second patch of ripples downstream of the patch under investigation (figure \ref{fig:schematic}b). For the presentation purpose, we assume that ripples in both patches have the same normalized wavenumber $n_b$=2 and amplitude $a_b$=4$\pi$/100. The distribution of energy on each patch, and in the area between the two patches, similar to the case of \S\ref{sec:wall} is strongly dependent on $L_m$ (the distance between the two patches). Distribution of energy density $\widetilde E$ for $L_m/\lambda_b$=0, 0.125, 0.250, 0.375 and 0.500 is shown in figure \ref{fig:two_patch_ener}a for two identical patches of $n$=4 seabed ripples. Note that the actual distance between the patch in each case is $q\lambda_b+L_m$ ($q$ positive integer) where in the case of figure \ref{fig:two_patch_ener}a, $q$=4. But as before $q$ does not play any role and it is $L_m$ that determines the energy distribution. Figure \ref{fig:two_patch_ener}a shows that for $L_m/\lambda_b$=0, 0.125 and 0.250 energy continuously increases over two patches and is constant in the area between the two patches. For $L_m/\lambda_b$=0.375 and 0.50, $\widetilde E$ increases over the first patch, but decreases over the second patch in such a way that it gains a maximum in the area between the two patches: that is, energy is trapped in this area. To see the behavior of the amplitude of each mode, we show in figures \ref{fig:two_patch_ener}b, c the spatial evolution of amplitudes of transmitted and reflected resonant modes (first five modes, i.e. modes 1, 3, 5, 7 and 9) over the two patches of ripples with $L_m/\lambda_b$=0.50. Similar to the energy density, amplitudes of all modes consistently increase over the first patch, and in a similar way decrease over the second patch. Interestingly, at the end of the second patch, all the energy is back to the original incident wave energy: an upstream/downstream observer sees absolutely no trace of the two seabed patches on the upstream/downstream wave field.

The behavior of energy is also a function of number of ripples in the patch as well as the number of ripples in the neighboring patch. For a total number of ripples in both patches equal to $n$=8, we show in figures \ref{fig:two_patch_ener}d-f how energy density changes at the beginning of the first patch $\widetilde E(0)$, in the middle of the two patches $\widetilde E(n_1\lambda_b)$, and at the end of the second patch $\widetilde E[(n_1+n_2+q)\lambda_b+L_m]$. In all cases, energy at the beginning of the first and at the end of the second patch obtains a global minimum for $L_m/\lambda_b$=0.5. For the area between the two patches, energy is maximum for $L_m/\lambda_b$=0.5. The energy density upstream and downstream of the two patch system is only a function of the total number of ripples and not a function of how they are distributed in the two patches.

\section{Concluding remarks}\label{sec:conc}

We presented here, analytically supported by direct simulation,  that the energy distribution of internal waves over a patch of seabed undulations can be strongly dependent upon the distance of the patch to the neighboring seabed features. Specifically, we considered two neighboring features: a second patch of seabed undulations and a vertical wall (a perfect reflector). We showed that accumulation of internal waves energy may be an order of magnitude larger or smaller at specific areas of a patch, solely based on where the neighboring feature is. The wall or the second patch, with right properties and placement, can also completely cancel the effect of the first patch in such a way that  upstream and downstream observers see no trace of the patch in their local wave field. The phenomenon elucidated here may influence, potentially significantly, the distribution of internal waves energy near steep oceanic ridges and continental slopes. 


\bibliographystyle{agufull08}
\bibliography{KA_stratified,Bragg3,Cloaking,MyRef,Mainbiblio,MyPapers}

\end{document}